\documentclass[nofootinbib,twocolumn,aps,pre,superscriptaddress,citeautoscript,floatfix, groupedaddress]{revtex4-1}
\usepackage{graphicx}
\usepackage{amsmath,amssymb}
\usepackage[dvipsnames]{xcolor}

\begin{document}
\title{Charge Oscillations in Ionic Liquids: A Microscopic Cluster Model}
\author{Yael Avni}
\author{Ram M. Adar}
\author{David Andelman}
\email{andelman@tauex.tau.ac.il}
\affiliation{Raymond and Beverly Sackler School of Physics and Astronomy,\\ Tel Aviv University, Ramat Aviv 69978, Tel Aviv, Israel}

\begin{abstract}
In spite of their enormous applications as alternative energy storage devices and lubricants, room temperature ionic liquids (ILs) still pose many challenges from a pure scientific view point. We develop an IL microscopic theory in terms of ionic clusters, which describes the IL behavior close to charged interfaces.
The full structure factor of finite-size clusters is considered and allows us to retain fine and essential details of the system as a whole.
Beside the reduction in the screening, it is shown that ionic clusters cause the charge density to oscillate near charged boundaries, with alternating ion-size thick layers, in agreement with experiments. We distinguish between short-range oscillations that persist for a few ionic layers close to the boundary, as opposed to long-range damped oscillations that hold throughout the bulk. The former can be captured by finite-size ion pairs, while the latter is associated with larger clusters with pronounced quadrupole (or higher) moment. The long-wavelength limit of our theory recovers the well-known Bazant-Storey-Kornyshev (BSK) equation in the linear regime, and elucidates the microscopic origin of the BSK phenomenological parameters.

\end{abstract}

\maketitle

Room-temperature ionic liquids (ILs) have been recently a subject of intense research due to their numerous applications as electrolytes in batteries, fuel cells and supercapacitors, and as molecular ``green" lubricants~\cite{Galiski2006, Buzzeo2004, Armand2011, supercapacitor, Lubrication1, Lubrication2}. In addition, being a highly interacting coulomb system, the statistical mechanics modelling of ILs, especially near electrified interfaces, continues to pose a great theoretical challenge~\cite{Kornyshev2007, Fedorov2014}.

As ILs are solvent-free electrolyte systems, they can be modelled as concentrated ionic solutions. Although ILs are frequently compared with dilute ionic solutions, for which the theoretical understanding and fits to experiments are well established~\cite{Safinya}, experiments and simulations reveal key differences between these two systems. Unlike dilute solutions, the IL charge distribution near charged interfaces is often non-monotonic and can decay in an oscillatory manner. A combination of x-ray structure measurements~\cite{Mezger2008}, force measurements~\cite{Hayes2010, Hoth2014, Li2013, Perkin2012} and molecular dynamics simulations~\cite{Kornyshev2008a, Kornyshev2008b, Fedorov2013} revealed that close to charged interfaces, cations and anions form alternating layers of about one ionic diameter in thickness. When confined between two charged surfaces, ILs can even go through a phase transition into a solid-like phase \cite{Bocquet}. Another distinct feature is {\it under-screening}. Namely, the screening in ILs and in concentrated ionic solutions is much weaker than the dilute-solution prediction~\cite{Smith2016,Underscreening2017, Israelachvili2013, Gebbie2017}.

The most common theory of ionic solutions is the Poisson-Boltzmann (PB) theory. This is a mean-field (MF) theory~\cite{Safinya} that is valid for low ionic concentrations. At high concentrations, as in the case of ILs, ionic correlations that are neglected in the PB theory become important~\cite{Netz2000Beyond, Naji2013}. Nevertheless, in the attempt to construct an effective theory for ILs, the PB equation has been often used with further modifications that were supposed to account for deviations from the dilute regime~\cite{Lauw2009, Gavish2016, Gavish2017, Girotto2017}. Other approaches include liquid-state theory~\cite{Tosi1986, Zerah1986} and one-dimensional lattice-gas models~\cite{1D2012a, 1D2012b}.

One of the significant contributions to describe theoretically ILs has been suggested by Bazant, Storey and Kornyshev (BSK)~\cite{Bazant}, who developed an equation that modifies the PB equation in two important ways. It uses a lattice-gas entropy instead of an ideal-gas one, and it has a biharmonic term in the electrostatic potential, $\nabla ^2 \nabla ^2 \psi(\bf r)$. While the entropy modification is more standard in incorporating steric interactions~\cite{Borukhov1997}, the additional biharmonic term is a purely phenomenological way to incorporate ionic correlations via a non-local dielectric constant~\cite{Kornyshev1978}, whereas other works suggested that it can be related to structural non-electrostatic interactions~\cite{Blossey2017}. The BSK equation requires fine-tuning of the model parameters in order to obtain the experimentally observed wavelength of charge-density oscillations~\cite{Mezger2008}, but it comes at the expense of too short decay length, which is not realized in physical systems. In this Letter, we address those issues  and offer a more microscopic approach to ILs, which predicts some of their main features.

It has long been conjectured that ionic clusters are likely to form in concentrated electrolytes~\cite{Bjerrum1926, Abascal1994, Levin1996}, due to the strong electrostatic interactions and reduced entropy. Recent simulations that follow single-ion trajectories in ILs support this description and show that the fraction of free ions is only around 20\%~\cite{Feng2018}. We note that ionic clusters ({\it e.g.}, ion pairs) in concentrated ionic solutions were investigated previously~\cite{VanRoij2009, LeeAlpha, Kjellander2016, Adar2017} and were shown to increase the screening length relative to the Debye length, suggesting under-screening~\cite{Israelachvili2013}. However, the connection between cluster formation and the observed charge oscillations near interfaces has not yet been fully explored.

In this Letter, we use an elegant yet simple cluster picture to obtain an effective MF theory for ILs. For small electric fields (or small charge density), we show that the clusters induce formation of layers near charged interfaces, having alternating charge and a microscopic thickness. While our theory is derived from a microscopic model with a clear interpretation of all the physical parameters, it is able to reproduce the qualitative mesoscopic features observed in experiments.

\begin{figure}
\includegraphics[width = 1.0\columnwidth,draft=false]{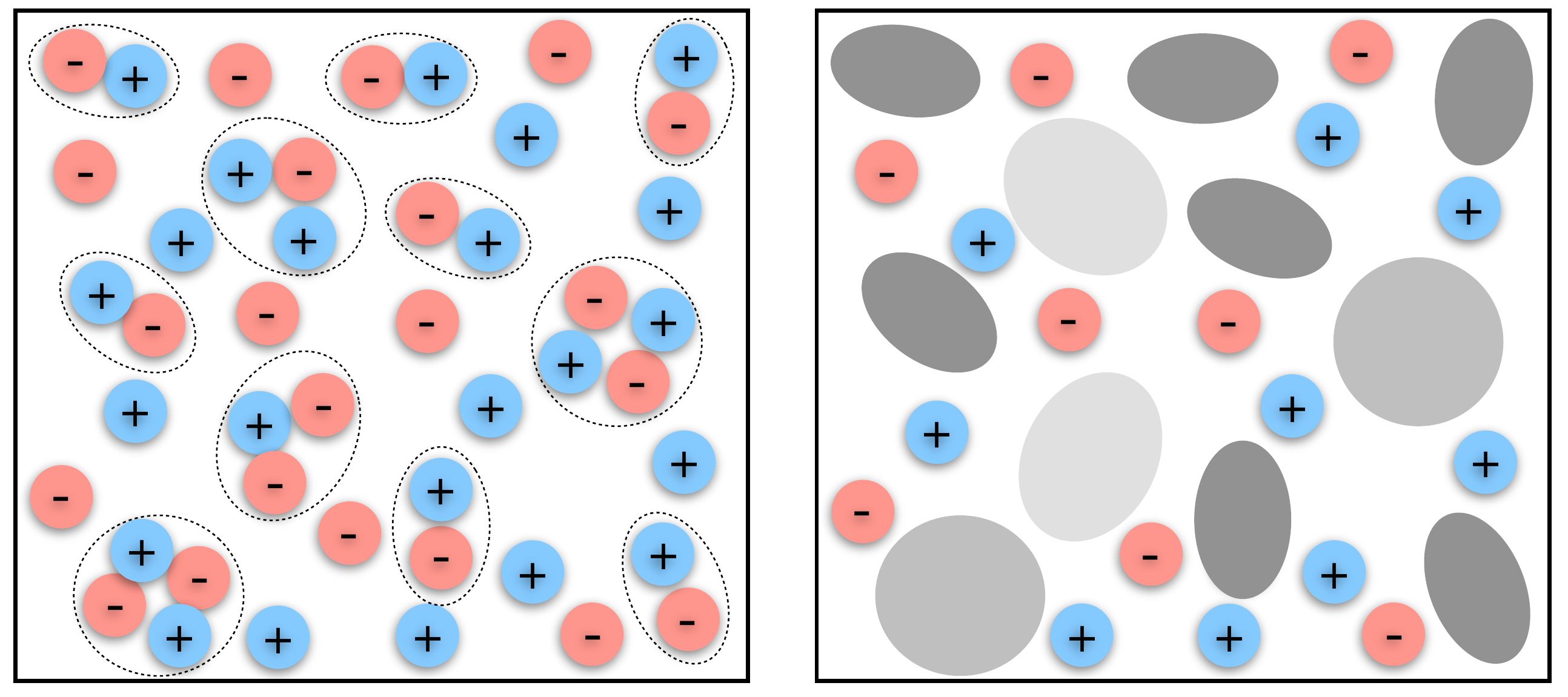}
\caption{\textsf{(color online)  A schematic drawing of ionic clusters in ILs. Densely packed free ions (left) can be viewed as a combination of free ions and clusters of bound ions. Clusters with the same ionic composition ({\it e.g.}, ion pairs) are treated in the same way (right).}}
\label{Fig1}
\end{figure}

{\it The cluster model ---} We consider a system of positive and negative ions of charge $\pm q$. The bulk concentration of the two types of ions is $n_0$ and the system is in thermal equilibrium at temperature $T$. As there is no additional solvent surrounding the ions, the relative dielectric constant is the same as the vacuum one, $\varepsilon=1$. The three characteristic length scales are the Bjerrum length, $l_{{\rm B}}=\beta q^{2}/4\pi\varepsilon_{0}$, where $\beta=1/k_{\rm B} T$ and $\varepsilon_0$ is the vacuum permittivity, the Debye length, $\lambda_{\rm D}=1/\kappa_{\rm D}=(8\pi l_{\rm B} n_0)^{-1/2}$, and the ionic size, $a$ (assumed for simplicity to be the same for both cations and anions). The ionic concentration can be expressed in units of $a^{-3}$ such that  $2n_0= \gamma a^{-3}$ where the fraction $\gamma$ is smaller than unity but a substantial fraction of it. We consider room-temperature ILs ($T\simeq 300\,$K) and  take $q=e$ (the electron charge), such that $l_{\rm B}\simeq 55\,$nm. The ion size in typical ILs is of the order $a \simeq 1\,$nm. For simplicity, we assume $\gamma \simeq 0.5$, leading to $\kappa_{\rm D} a= \sqrt{4\pi \gamma}\sqrt{l_{\rm B}/a}\simeq 20\gg1$. Such large $\kappa_{\rm D}$ values indicate that the regular PB theory is no longer valid.

We postulate the formation of various types of clusters, as a way to incorporate ionic correlations. Each ion cluster has a specific internal configuration, defined up to rotation and translation of the entire cluster (Fig.~\ref{Fig1}). We treat all clusters of the same ionic composition on equal footing, and neglect the steric repulsion between clusters. The latter can be justified in the linear regime, as will be employed further below. The resulting MF equation~\cite{Levy2013, Buyukdagli2013, Frydel2013} can be written as

\begin{equation} \label{MF}
\begin{split}\nabla^{2}\psi\left({\bf r}\right)=& \\
 -&\frac{1}{\varepsilon_{0}}\bigg[\rho_{f}\left({\bf r}\right)+\sum_{m}n_{m}\int{\rm d}\Theta\int{\rm d}{\bf r}''\rho_{m}\left(\Theta;{\bf r}-{\bf r}''\right)\\
\times & \exp\left(-\beta\int{\rm d}{\bf r}'\rho_{m}\left(\Theta;{\bf r}'-{\bf r}''\right)\psi\left({\bf r}'\right)\right)\bigg],
\end{split}
\end{equation}
where $\psi(\bf r)$ is the electrostatic potential, $n_m$ and $\rho_m$ are the bulk concentration and internal charge density of the $m$-type cluster, respectively, and $\rho_f$ is some fixed charge density. The cluster charge-density depends on the orientation, $\Theta$, defined by the three Euler angles, $\varphi$, $\theta$ and $\zeta$, with ${\rm d}\Theta=\frac{1}{8\pi^{2}}\sin\theta\,{\rm d}\varphi\,{\rm d}\theta\,{\rm d}\zeta$. The sum in Eq.~(\ref{MF}) is over all cluster types, including the free ions. The Boltzmann factor for each cluster depends on the value of the electrostatic potential $\psi(\bf r)$ within the cluster volume, rendering the equation non-local. The bulk cluster concentrations, $n_m$, depend on various physical conditions~\cite{Levin1996,Yokoyama1975, Ebeling1980}. We treat the concentrations as model parameters and relate them below to measurable IL properties.

We now take the limit of small electrostatic potentials, $\beta q \psi({\bf r})\ll1$, and linearize Eq.~(\ref{MF}). Within this linear limit, steric interactions are expected to have negligible effect on electrostatic properties~\cite{Borukhov1997}. In Fourier space, the linearized version of Eq.~(\ref{MF}) takes the form
\begin{equation} \label{FT}
-k^{2}\tilde{\psi}_{{\bf k}}=\frac{1}{\varepsilon_{0}}\bigg[\beta\sum_{m}n_{m}S_{m}\left(k\right)\tilde{\psi}_{{\bf k}}\,-\,\tilde{\rho}_{f,{\bf k}}\bigg],
\end{equation}
where $\tilde{f}_{{\bf k}}=\int{\rm d}{\bf r}f({\bf r}){\rm e}^{i{\bf k}\cdot{\bf r}}$ is the Fourier transform of $f({\bf r})$ and $S_m(k)$ is the charge density structure-factor of the $m$-cluster, defined by
\begin{equation} \label{Structure}
{S_{m}\left(k\right)\equiv\int{\rm d}\Theta\left|\tilde{\rho}_{m,{\bf k}}\left(\Theta\right)\right|^{2}=\frac{1}{4\pi}\int{\rm d}\Omega_{{\bf k}}\left|\tilde{\rho}_{m,{\bf k}}\right|^{2}}.
\end{equation}
The last equality in Eq.~(\ref{Structure}) is due to the observation that the average over the orientation angles of the cluster can be replaced by an average over $\Omega_{\bf k}$, the solid angle in $\bf k$-space.

In order to study the IL behavior near an electrified boundary, we consider in Eq.~(\ref{FT}) the simple case of a charged surface, with fixed charge density, $\sigma_0$, immersed in the liquid at $z=0$ such that $\rho_{f}\left({\bf r}\right)=\sigma_0\delta\left(z\right)$. More realistic surfaces, {\it i.e}, a thick dielectric or conductor, present a non-trivial boundary condition problem. In such cases, the exact surface properties need to be specified, for example via a surface free energy \cite{Podgornik2018}.

\begin{figure}
\includegraphics[width = 1\columnwidth,draft=false]{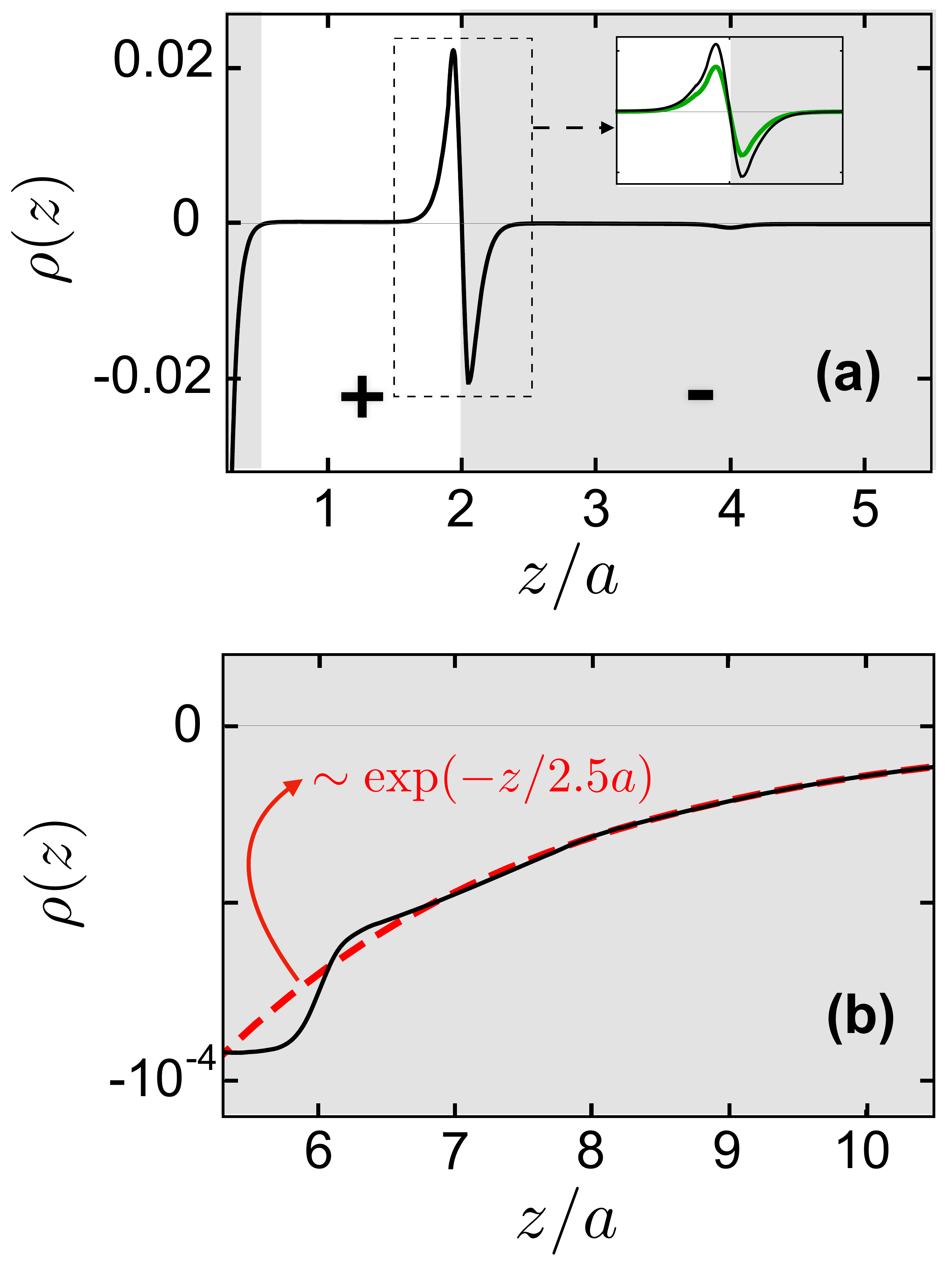}
\caption{\textsf{(color online) The charge density, $\rho(z)$, rescaled by $\sigma_0/a$, as a function of the distance from a charged surface for (a) short and (b) long  distances. Only free ions and pair clusters are considered, with corresponding fractions: $\alpha_1 = 0.1$ and $\alpha_p=0.9$ and $\kappa_{\rm D} a= 20$. The solid black line in (a) and (b) is the exact solution from Eq.~(\ref{poles}). The green line in the inset of (a) is plotted for $\alpha_1 = 0.4$ and $\alpha_p=0.6$, for comparison. The dashed red line in (b) is a fitted exponential function with the decay length $\kappa_1^{-1}$, taken from the first pole of Eq.~(\ref{poles}). White and gray backgrounds distinguish between positive and negative charge densities, respectively. The ion-pair clusters reverse the sign of $\rho(z)$ at short distances, yet in the long-distance limit, $\rho(z)$ has a pure monotonic decay.}}
\label{Fig2a}
\end{figure}
\begin{figure}
\includegraphics[width = 1\columnwidth,draft=false]{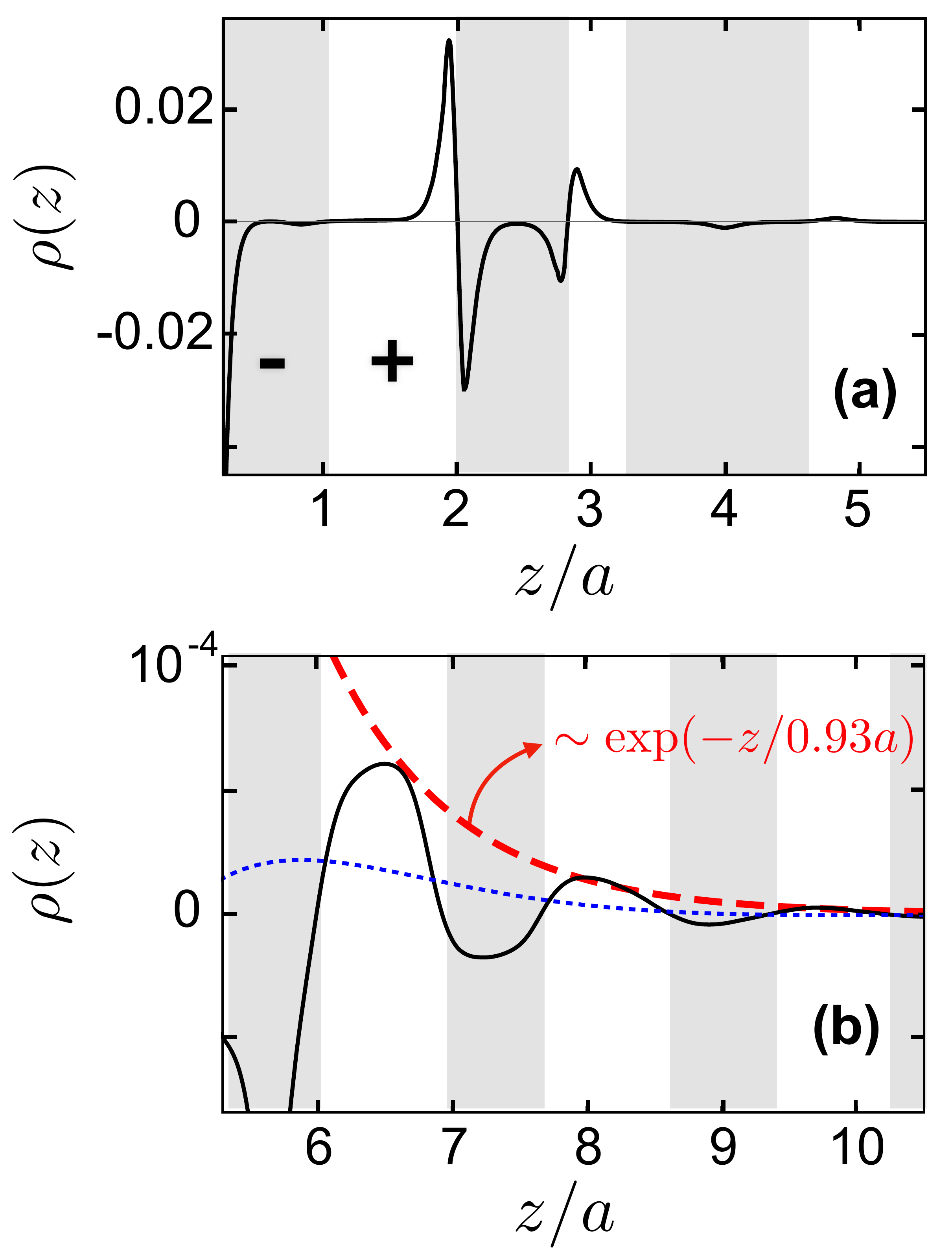}
\caption{\textsf{(color online) The same plot as in Fig.~\ref{Fig2a}, with the addition of {\it square} clusters on top of free ions and pair clusters, with fractions: $\alpha_1 = 0.3$, $\alpha_p=0.1$ and $\alpha_s=0.6$ and $\kappa_{\rm D} a= 20$. The solid black line is the exact solution from Eq.~(\ref{poles}). In (b), the dashed red line is a fitted exponential function with the decay length $\kappa_1^{-1}$, taken from the first pole of Eq.~(\ref{poles}), and the dotted blue line is the approximated solution, corresponding to Eq.~(\ref{BSK}). Due to the presence of square clusters, the charge density oscillates even at large distances. The disagreement between the exact and approximated solutions shows that high-order terms in the expansion of $S_m(k)$ in powers of $k$ are essential to capture the correct oscillation wavelength.}}
\label{Fig2b}
\end{figure}
Denoting $\alpha_m$ as the fraction of ions that belong to the $m$-type cluster, we get that for an $m$-cluster with $N_m$ ions, $n_{m}=2n_{0}\alpha_{m}/N_{m}$ and $\sum_{m}\alpha_{m}=1$. Introducing the normalized quantities: $\hat{z}=z/a$, $\hat{k}=ka$ and ${\tilde{S}_{m}(\hat{k})=S_{m}(\hat{k})/q^{2}N_{m}}$, we obtain
\begin{equation}\label{poles}
\psi\left(\hat{z}\right)=\frac{\sigma_0 a}{\varepsilon_{0}}\frac{1}{2\pi}\int {\rm d}\hat{k}\,\,\frac{{\rm e}^{-i\hat{k}\hat{z}}}{\hat{k}^{2}+\kappa_{D}^{2}a^{2}\sum\limits _{m}\alpha_{m}\tilde{S}_{m}(\hat{k})}.
\end{equation}
The denominator of Eq.~(\ref{poles}) can be interpreted in terms of a {\it non-local dielectric constant}~\cite{Bopp1996,  Buyukdagli2013, Buyukdagli2013_b, Buyukdagli2014, Kjellander2016, Kjellander2018, Budkov2018}.

The value of the $k$-integration in Eq.~(\ref{poles}) is determined from the residue theorem by the poles in the lower half of the complex plane, which are given by the equation ${\hat{k}^{2}+\kappa_{D}^{2}a^{2}\sum_{m}\alpha_{m}\tilde{S}_{m}(\hat{k})=0}$ (see also Ref.~\cite{Adar2019}). Each pole can be written as $k=\pm\omega-i\kappa$ with positive $\kappa,\omega>0$. The contribution of each pole to $\psi(z)$ is $A {\rm e}^{-\kappa z}\cos\left(\omega z+\phi \right)$ where $A$ and $\phi$ are real constants. Since $\tilde{S}_{m}(\hat{k})$ is a non-negative function, the poles will always have a non-zero imaginary value, ${\kappa \neq 0}$, causing $\psi(z)$ to decay exponentially. However, some poles might have a non-zero real value as well, ${\omega \neq 0}$, corresponding to oscillating decaying modes. The most dominant pole at large $z$ distances (``the first pole") has the smallest imaginary part (denoted by $\kappa_1$). Therefore, the condition for long-range damped oscillations, as opposed to a purely long-range decay, is that the first pole is not purely imaginary. We show below how this is determined by the cluster composition.

The exact clusters that are likely to form in ILs depend on the molecular structure of the cations and anions. We implement the general formalism on a simplified system that contains only two types of clusters, other than free ions: an {\it ionic pair}, composed of a cation and an anion with internal charge density ${\rho\left({\bf r}\right)=q\left[\delta\left({\bf r}-a\hat{x}\right)-\delta\left({\bf r}+a\hat{x}\right)\right]}$, and an {\it ionic square} cluster, composed of two anti-parallel ionic pairs with $\rho\left({\bf r}\right)=q[\delta\left({\bf r}-a\hat{x}-a\hat{y}\right)+\delta\left({\bf r}+a\hat{x}+a\hat{y}\right)-\delta\left({\bf r}-a\hat{x}+a\hat{y}\right)-\delta\left({\bf r}+a\hat{x}-a\hat{y}\right)]$. We denote $m=1$ for the single positive/negative ion cluster, $m=p$ for the pair cluster and $m=s$ for the square cluster, resulting in
\begin{equation}
\begin{split}
\tilde{S}_{1}(\hat{k})& =1\\
\tilde{S}_{p}(\hat{k})& =1-\frac{\sin(2\hat{k})}{2\hat{k}}\\
\tilde{S}_{s}(\hat{k})& =1-\frac{4 \sin(2\hat{k})-\sqrt{2}\sin(2\sqrt{2}\hat{k})}{4\hat{k}}.
\end{split}
\end{equation}
Substituting the above structure factors in Eq.~(\ref{poles}), the electrostatic potential profile $\psi(z)$ and the charge density, ${\rho\left(z\right)=-\varepsilon_{0}\psi''(z)}$, are obtained.

{\it Results and discussion ---} In Figs.~\ref{Fig2a} and \ref{Fig2b}, we show the charge density as a function of the distance from a charged surface, for different choices of the fractions $\alpha_1$, $\alpha_p$ and $\alpha_s$. We focus on the charge density at distances larger than $a/2$, because for smaller distances,  short-range interactions between the clusters and surface should also be taken into account. This can be done in future studies, {\it e.g.}, by limiting the spatial configurations of the clusters.

Our results show that if $\alpha_s=0$, {\it i.e}, only free ions and ion pairs are present, the first pole of Eq.~(\ref{poles}) is always purely imaginary, leading to a monotonic decay of $\rho(z)$ at long distances (Fig.~\ref{Fig2a}b). However, at short distances (Fig.~\ref{Fig2a}a), the complex higher-order poles are dominant and cause the charge density to reverse its sign, corresponding to short-range oscillations. These short-range oscillations exist due to the finite distance between the ions inside the pair, and they vanish for point-like pairs. Moreover, as the ratio $\alpha_p/\alpha_1$ increases, the short-range oscillations become more pronounced (see inset in Fig.~\ref{Fig2a}a). The case of $\alpha_s=0$ resembles the model considered by Buyukdagli {\it et. al.}~\cite{Buyukdagli2013, Buyukdagli2014} of point-like ions embedded in a polar liquid with a finite solvent size.

When square clusters are present as well (Fig.~\ref{Fig2b}), we find that above some concentration threshold (discussed below), the first pole has a non-zero real part. This leads to damped charge-density oscillations, both close to the surface and in the distal region. In fact, the long-range behavior in this case is dominated not by a single pole, but by two poles that have very close imaginary values yet different real values (see supplemental material~\cite{Supplemental}). The wavelength of both short- and long-range oscillations is of the order of the ionic size (see Figs.~\ref{Fig2a} and \ref{Fig2b}), in agreement with experiments and simulations~\cite{Mezger2008, Hoth2014, Kornyshev2008a}. We have thus shown that the formation of ionic clusters, when taken into account properly (with a finite separation between the ions inside each cluster), leads to the formation of IL layers with alternating charge.
We note that these charge oscillations are different than the structural density oscillations that are common in both charged and uncharged liquids~\cite{Israelachvili1988}.

We turn now to the effective screening length (playing the role of the Debye length $\lambda_{\rm D}$), determined by the decay length of the electrostatic potential, $\psi(z)$. At short distances, the decay length is not defined, because $\psi(z)$ is constructed from multiple modes, each with a different characteristic decay length. In the large distance limit, the effective screening length is defined by the inverse of the imaginary part of the first pole, $\kappa_1^{-1}$. From this definition, we find that the screening length is of the order of the ionic size $a$ (see dashed red lines in Figs.~\ref{Fig2a}b and~\ref{Fig2b}b), which is much larger than the decay length, ${\lambda_{\rm D}=\kappa_{\rm D}^{-1}\approx0.05a}$, predicted by the Debye-H\"uckel theory for dilute ionic solutions. Our results, thus, qualitatively reproduce the under-screening effect. This is of no surprise since the cluster picture initially assumes that only a fraction of the total number of ions are free, and participate in the screening. We note that experiments suggested an even larger screening length~\cite{Gebbie2017}.

\begin{figure}
\includegraphics[width = 1 \columnwidth,draft=false]{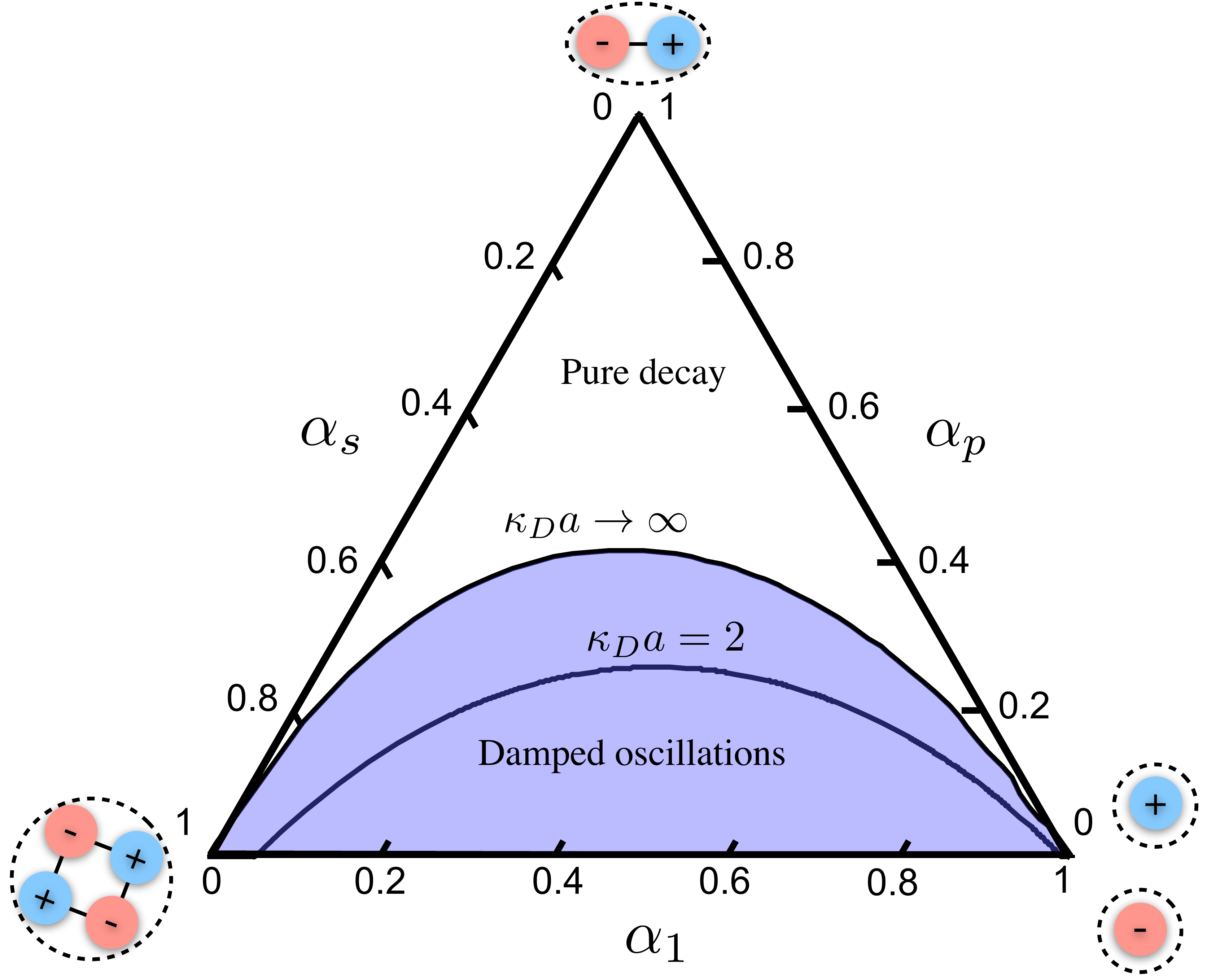}
\caption{\textsf{(color online) Ternary phase diagram for $\kappa_{\rm D} a=2$ and $\kappa_{\rm D} a\to\infty$. Depending on the cluster fractions $\alpha_1$, $\alpha_p$ and $\alpha_s$, $\rho(z)$ either has a regular decay (above the curves) or damped oscillations  (below the curves)}. The diagram is restricted to the large distance limit.}
\label{Fig3}
\end{figure}

The transition from pure decay to an oscillatory one at large distances is depicted in Fig.~\ref{Fig3}, where the phase diagram as a function of the fractions $\alpha_1$, $\alpha_p$ and $\alpha_s$ is shown. As $\alpha_1+\alpha_p+\alpha_s=1$, a ternary phase diagram is plotted in Fig.~\ref{Fig3}. Although it is hard to see from the diagram, the vertex $\alpha_1=1$ always lies in the region of a pure monotonic decay, which means that $m\neq 1$ clusters are needed for damped oscillation to occur. The diagram shows that the long-range charge oscillations are enhanced by both the free ions and square clusters, but are suppressed by the pair clusters. This picture is not limited to our specific choice of clusters. Clusters with a pronounced dipole moment favor a longe-range monotonic decay of the charge density, while clusters with small dipole moment but pronounced quadrupole or higher multipoles, are likely to cause long-range oscillations. Figure~\ref{Fig3} shows that the area in phase space of long-range oscillation grows when $\kappa_{\rm D} a$ increases, until it reaches a limiting value for $\kappa_{\rm D} a \to \infty$. In interpreting Fig.~\ref{Fig3}, one should keep in mind that the fractions $\alpha_1$, $\alpha_p$ and $\alpha_s$ depend, in general, on $\kappa_{\rm D} a$ as well.

{\it Relation to the BSK equation ---} Our cluster theory can be related to the BSK equation~\cite{Bazant}. By expanding $S_m(k)$ to 4th order in $k$, corresponding to a long-wavelength approximation and transforming Eq.~(\ref{FT}) back to real space, we get an expression similar to the the linearized BSK equation,
\begin{equation} \label{BSK}
\Big[\nabla^{2}-\xi^{2}\nabla^{2}\nabla^{2}-\kappa_{\rm eff}^{2}\Big]\psi\left({\bf r}\right)=-\frac{\rho_{f}\left({\bf r}\right)}{\varepsilon_{0}\varepsilon_{\rm eff}},
\end{equation}
where $\varepsilon_{\rm eff}$, $\kappa_{\rm eff}$ and $\xi$ are given by:
\begin{equation} \label{constants1}
\begin{split}
\varepsilon_{\rm eff} &=1+\frac{\beta}{2\varepsilon_{0}}\sum_{m}n_{m}S_{m}''\left(0\right)\\
\kappa_{\rm eff}^{2} &=\frac{\beta}{\varepsilon_{0}\varepsilon_{\rm eff}}\sum_{m}n_{m}S_{m}\left(0\right)\\
\xi^{2} &=\frac{\beta}{24\varepsilon_{0}\varepsilon_{\rm eff}}\sum_{m}n_{m}S_{m}^{(4)}\left(0\right).
\end{split}
\end{equation}
An equation similar to Eq.~(\ref{BSK}) was derived for ions dissolved in quadrupolar dielectrics~\cite{Slavchov1,Slavchov2}. However, in that case, it is the solvent that is quadrupolar and is distributed homogeneously in space, unlike the clusters in our model.

The derivatives of the structure factor and, consequently, the above coefficients, are related to different multipole moments of the cluster charge-density~\cite{Supplemental}. For $\xi^2<0$, higher orders of $S_m(k)$ must be taken into account. In the specific example used here, the free ions ($\alpha_1$), ionic pairs ($\alpha_p$) and ionic squares ($\alpha_s$), give
\begin{equation} \label{constants1}
\begin{split}
\varepsilon_{{\rm eff}} & =1+\frac{2\alpha_{p}\kappa_{\rm D}^{2}a^{2}}{3}\\
\kappa_{{\rm eff}}^{2} &=\frac{\alpha_{1}\kappa_{\rm D}^{2}}{\varepsilon_{{\rm eff}}}\\
\xi^{2} & =\frac{2\kappa_{\rm D}^{2}a^{4}\left(2\alpha_{s}-\alpha_{p}\right)}{15\varepsilon_{{\rm eff}}}.
\end{split}
\end{equation}
This relates the fractions $\alpha_1$, $\alpha_p$ and $\alpha_s$ to mesoscopic quantities. Note that the parameter $\xi$ is equivalent to the phenomenological `correlation length' in the BSK theory~\cite{Bazant}, denoted there by $l_c$, with the difference that here it is given in terms of the cluster properties.

The linearized BSK equation predicts damped long-range oscillation only if $\xi^2 \kappa_{\rm eff}^2>1/4$. In the limit $\kappa_{\rm D} a \gg 1$, this condition becomes
\begin{equation} \label{condition}
\frac{3\alpha_{1}\left(2\alpha_{s}-\alpha_{p}\right)}{10\alpha_{p}^{2}}>\frac{1}{4},
\end{equation}
predicting long-range oscillation for large $\alpha_1$ and $\alpha_s$ and small $\alpha_p$, similar to the predictions of Eq.~(\ref{poles}).
In fact, if we construct the diagram in Fig.~\ref{Fig3} from the approximated Eq.~(\ref{BSK}), its qualitative features are unchanged~\cite{Supplemental}, meaning that the long-wavelength approximation captures the transition from a monotonic decay to damped charge oscillations in the limit $z / a \gg 1$.

However, the approximation incorrectly predicts that the wavelength of the long-ranged oscillations is much larger than the ionic size, as shown by the dashed blue line of Fig.~\ref{Fig2b}b. This should not be of a surprise as short-wavelength phenomena are captured by large wave-numbers (large $k$'s). It is, therefore, crucial to consider high-order terms in $k$, in order to obtain the layering observed in experiments~\cite{Mezger2008}.

To summarize, we developed a microscopic model for ionic liquids that is based on the assumption that a substantial fraction of the ions aggregates into clusters, as is indicated by recent simulations. Our theory predicts both the under-screening effect and interfacial charge-density oscillations of approximately one ionic layer thickness. This is in accord with experiments and simulations. We show that ionic pairs lead to short-range charge-oscillations with an amplitude that increases with the pairs fraction, and that more complex clusters can lead to long-ranged charge oscillations.

Our theory is limited to the linear regime that can be justified for relatively small charge densities. In order to go beyond the latter approximation, steric interactions must be included in the theory. The advantage of the clusters description is in its simplicity and clear interpretation, which quite remarkably capture key features of ionic liquids.

\bigskip\bigskip
{\em Acknowledgements}~~
We would like to thank D. Harries, A. Kornyshev, S. Perkin, K. Pivnic, R. Podgornik, Y. Tsori, and M. Urbakh for fruitful discussions, and R.~Blossey and S. Buyukdagli for useful suggestions. This work was supported by the Israel Science Foundation (ISF) under grant no. 213/19.

\onecolumngrid
\newpage

\appendix
\section*{Appendix A: Expressions for $\kappa_{\rm eff}$, $\varepsilon_{\rm eff}$ and $\xi$ \\in terms of multipole moments of the cluster charge density}
The parameters $\kappa_{\rm eff}$, $\varepsilon_{\rm eff}$ and $\xi$ in Eq.~(7) in the letter we need to distinguish somehow between equations in the letter and in the supplemental material are related to the structure factor and its derivatives, $S_m(0)$, $S''_m(0)$ and $S^{(4)}_m(k)$, and can be written in terms of the multipole moments of the cluster charge density.

The charge density of an $m$-type cluster at position ${\bf r}$ can be written as a Taylor expansion in the following way,
\begin{equation} \label{expansion}
\begin{split}
\rho_{m}\left({\bf r}\right) & =\int{\rm d}{\bf r}'\rho_{m}\left({\bf r}'\right)\delta\left({\bf r}-{\bf r}'\right)\\
 & \approx\int{\rm d}{\bf r}'\rho_{m}\left({\bf r}'\right)\bigg(\delta\left({\bf r}\right)-[\partial_{i}\delta\left({\bf r}\right)]{\bf r}'_{i}+\frac{1}{2}[\partial_{i}\partial_{j}\delta\left({\bf r}\right)]{\bf r}'_{i}{\bf r}'_{j}\\
 & -\frac{1}{3!}[\partial_{i}\partial_{j}\partial_{k}\delta\left({\bf r}\right)]{\bf r}'_{i}{\bf r}'_{j}{\bf r}'_{k}+\frac{1}{4!}[\partial_{i}\partial_{j}\partial_{k}\partial_{l}\delta\left({\bf r}\right)]{\bf r}'_{i}{\bf r}'_{j}{\bf r}'_{k}{\bf r}'_{l}\,+\,\mathcal{O}\left(r'^{5}\right)
\bigg).
\end{split}
\end{equation}
We define $q_{m}^{{\rm tot}} \equiv \int{\rm d}{\bf r}\,\rho\left({\bf r}\right)$, $p_{m}^{i} \equiv\int{\rm d}{\bf r}\,\rho\left({\bf r}\right){\bf r}_{i}$, $Q_{m}^{ij} \equiv\int{\rm d}{\bf r}\,\rho\left({\bf r}\right){\bf r}_{i}{\bf r}_{j}$, $H_{m}^{ijk} \equiv\int{\rm d}{\bf r}\,\rho\left({\bf r}\right){\bf r}_{i}{\bf r}_{j}{\bf r}_{k}$,
$G_{m}^{ijkl} \equiv\int{\rm d}{\bf r}\,\rho\left({\bf r}\right){\bf r}_{i}{\bf r}_{j}{\bf r}_{k}{\bf r}_{l}$, to be the monopole, dipole, quadrupole, octupole and hexadecapole moments of $\rho(\bf r)$, respectively. Equation~(\ref{expansion}) than becomes,
\begin{equation}
\rho_{m}\left({\bf r}\right) =q_{m}^{{\rm tot}}\delta\left({\bf r}\right)-p_{m}^{i}\partial_{i}\delta\left({\bf r}\right)+\frac{1}{2}Q_{m}^{ij}\partial_{i}\partial_{j}\delta\left({\bf r}\right)
 -\frac{1}{3!}H_{m}^{ijk}\partial_{i}\partial_{j}\partial_{k}\delta\left({\bf r}\right)+\frac{1}{4!}G_{m}^{ijkl}\partial_{i}\partial_{j}\partial_{k}\partial_{l}\delta\left({\bf r}\right)+....\,\,,
\end{equation}
and the Fourier transformed charge density, $\tilde{\rho}_{m,{\bf k}}=\int{\rm d}{\bf r}\,\rho_{m}\left({\bf r}\right){\rm e}^{i{\bf k}\cdot{\bf r}}$, is
\begin{equation} \label{rho_Fourier}
\begin{split}
\tilde{\rho}_{m,{\bf k}} & =q_{m}^{{\rm tot}}-ik_{i}{\bf p}_{m}^{i}-\frac{1}{2}k_{i}k_{j}Q_{m}^{ij} +\frac{1}{3!}ik_{i}k_{j}k_{k}H_{m}^{ijk}+\frac{1}{4!}k_{i}k_{j}k_{k}k_{l}G_{m}^{ijkl}+\mathcal{O}\left(k^{5}\right)
\,\,.
\end{split}
\end{equation}
We can calculate the charge-density structure-factor,
$S_{m}\left(k\right)=\frac{1}{4\pi}\int{\rm d}\Omega_{{\bf k}}\left|\tilde{\rho}_{m,{\bf k}}\right|^{2}$ ($\Omega_{\bf k}$ being the solid angle in ${\bf k}$-space), by substituting $\tilde{\rho}_{m,{\bf k}}$ of Eq.~(\ref{rho_Fourier});
\begin{equation}
\begin{split}
S_{m}\left(0\right) & =\big(q_{m}^{{\rm tot}}\big)^{2}\\
S''_{m}\left(0\right) & =\frac{1}{2\pi}\left({\bf p}_{m}^{i}{\bf p}_{m}^{j}-Q_{m}^{ij}q_{m}^{{\rm tot}}\right)\int{\rm d}\Omega_{{\bf k}}\frac{k_{i}k_{j}}{k^{2}}\\
S_{m}^{\left(4\right)}\left(0\right) & =\frac{1}{2\pi}\left(q_{m}^{{\rm tot}}G_{m}^{ijkl}-4{\bf p}_{m}^{i}H_{m}^{jkl}+3Q_{m}^{ij}Q_{m}^{kl}\right)\int{\rm d}\Omega_{{\bf k}}\frac{k_{i}k_{j}k_{k}k_{l}}{k^{4}}.
\end{split}
\end{equation}
Finally, using the identities,
\begin{equation} \label{S(0)}
\begin{split}
\frac{1}{4\pi}\int{\rm d}\Omega_{{\bf k}}\frac{k_{i}k_{j}}{k^{2}} & =\frac{\delta_{ij}}{3}\\
\frac{1}{4\pi}\int{\rm d}\Omega_{{\bf k}}\frac{k_{i}k_{j}k_{k}k_{l}}{k^{4}} & =\frac{1}{15}\left(\delta_{ij}\delta_{kl}+\delta_{ik}\delta_{jl}+\delta_{il}\delta_{jk}\right)
\end{split}
 \end{equation}
and the fact that $Q_{m}^{ij}$, $H_{m}^{ijk}$ and $G_{m}^{ijkl}$ are invariant under exchange of their indices, we obtain,
\begin{equation} \label{S(0)}
\begin{split}
\ensuremath{\kappa_{{\rm eff}}^{2} & =\frac{\beta}{\varepsilon_{0}\varepsilon_{{\rm eff}}}\sum_{m}n_{m}\left(q_{m}^{\text{tot}}\right)^{2}}\\
\varepsilon_{{\rm eff}} & =1+\frac{\beta}{3\varepsilon_{0}}\sum_{m}n_{m}\left[p_{m}^{2}-q_{m}^{\text{tot}}{\rm Tr}\left(Q_{m}\right)\right]\\
\xi^{2} & =\frac{\beta}{60\varepsilon_{0}\varepsilon_{{\rm eff}}}\sum_{m}n_{m}\bigg[\text{Tr}\left(Q_{m}\right)^{2}+2\text{Tr}\left(Q_{m}^{2}\right)-4 H_{m}^{iij}p_{m}^{j}+q_{m}^{\text{tot}}G_{m}^{iijj}\bigg].
\end{split}
 \end{equation}
From the above results, one can see that $\kappa_{\rm eff}$ increases only from contribution of clusters with a monopolar charge, $\varepsilon_{\rm eff}$ increases by clusters with large dipole moments but can be reduced by clusters that have both a monopole and quadrupole moments, and $\xi$, which is equivalent to the `correlation term' in the BSK theory is affected by higher-order multipoles. For a system that consists only of ideal multipoles, $\xi$ would be proportional to the quadrupolar concentration.

\section*{Appendix B: Comparison between full Eq.~(4) and the approximated Eq.~(6)}
\begin{figure}
\includegraphics[width = 0.6 \columnwidth,draft=false]{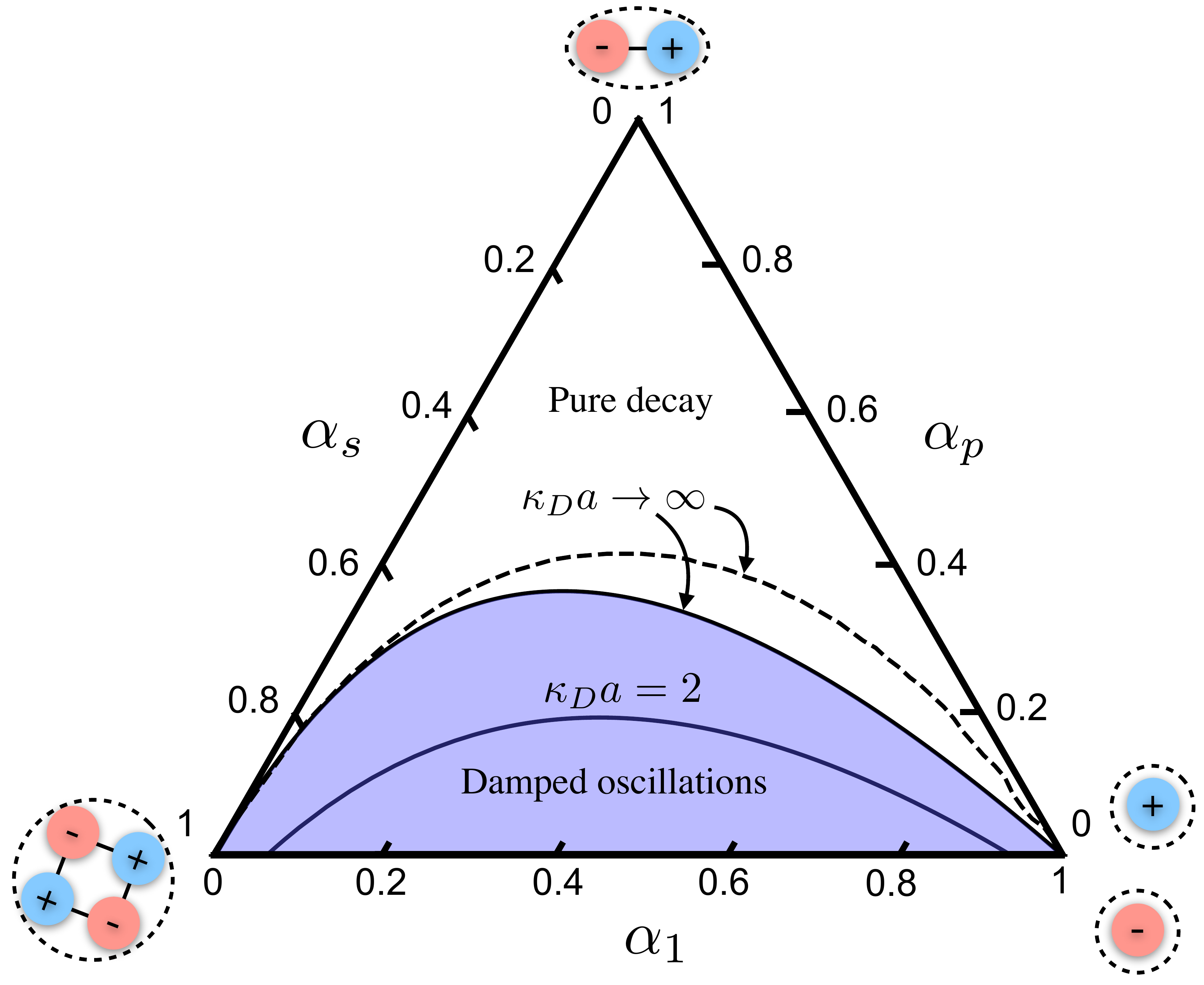}
\caption{\textsf{(color online) Ternary phase diagram of the liquid derived from the approximated Eq.~(6). To be compared with Fig.~4 in the Letter. The limit $\kappa_{\rm D} a \to \infty$ of the exact equation is plotted in a black dashed line for comparison.}}
\label{Fig1Supp}
\end{figure}

Figure~\ref{Fig1Supp} shows the phase diagram of the same system described in the Letter, which is composed of free ions, in-pair clusters and square clusters, only that here it is constructed by the approximated Eq.~(6).
The phase diagram of the approximated equation is very similar to the exact phase diagram (Fig.~4 in the Letter). This means that the long wavelength approximation captures the transition to long-range oscillations.

\begin{figure}
\includegraphics[width = 0.8 \columnwidth,draft=false]{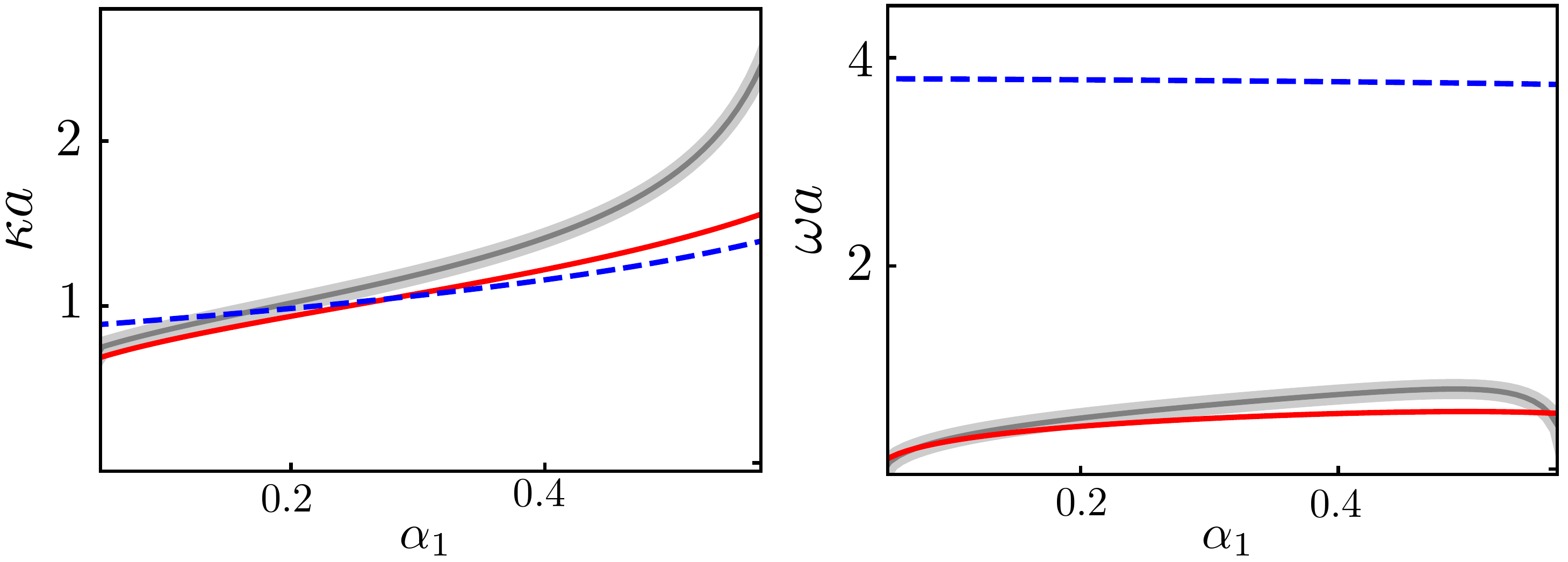}
\caption{\textsf{(color online) The imaginary (left) and real (right) parts of the first two poles of Eq.~(4) (full red and dashed blue lines), and the single pole  of Eq.~(6) (thick gray line), as a function of the fraction of free ions, $\alpha_1$. Other parameter values are $\alpha_p =0.25$ and $\kappa_{\rm D} a=20$. The poles are plotted only in the range of $\alpha_1$ in which there are long-range charge oscillations. The exact Eq.~(4) has two dominant poles, but the approximated Eq.~(6) captures only the pole with the long wavelength, and misses the short oscillations that capture the alternating layers in ionic liquids.}} 
\label{Fig2Supp}
\end{figure}

However, there is a large mismatch between the long-range oscillation wavelength of the full equation [Eq.~(4)] and the approximated equation [Eq.~(6)]. This can be understood from Fig.~\ref{Fig2Supp} presented here. The first two poles of Eq.~(4) have very close imaginary parts. Therefore, both contribute even at large distances of order $z/a\sim10$ (the exact distance at which one pole takes over the other depends on the specific choice of the cluster fractions $\alpha_i$). However, their real parts are very different. The approximated Eq.~(6) captures only the pole with the smaller real value, which corresponds to long wavelength oscillations, and it misses the short wavelengths.

\begin{thebibliography}{99}

\bibitem{Buzzeo2004}
M. C. Buzzeo, R. G. Evans, and R. G. Compton, Chem. Phys. Chem. {\bf 5}, 1106 (2004).

\bibitem{Galiski2006}
M. Gali{\'n}ski, A. Lewandowski, and I. St{\c e}pniak, Electrochim. Acta {\bf 51}, 5567 (2006).

\bibitem{Armand2011}
M. Armand, F. Endres, D. R. MacFarlane, H. Ohno, and B. Scrosati, in: {\it Materials for Sustainable Energy}, ed. by V. Dusastre (Nature publishing group and World Scientific, London, 2011), pp. 129-137.

\bibitem{supercapacitor}
M. Salanne, in: {\it Ionic Liquids II}, ed. by B. Kirchner and E. Perlt (Springer, Switzerland, 2017), pp. 29-53.

\bibitem{Lubrication1}
M. Palacio and B. Bhushan. Tribol. Lett. {\bf 40}, 247 (2010).

\bibitem{Lubrication2}
O. Y. Fajardo, F. Bresme, A. A. Kornyshev, and M. Urbakh, Sci. Rep. {\bf 5}, 7698 (2015).

\bibitem{Kornyshev2007}
A. A. Kornyshev, J. Phys. Chem. B. {\bf 111}, 5545 (2007).

\bibitem{Fedorov2014}
M. V. Fedorov and A. A. Kornyshev, Chem. Rev. {\bf 114}, 2978 (2014).

\bibitem{Safinya}
T. Markovich, D. Andelman, and R. Podgornik, in {\em Handbook of Lipid Membranes}, edited by C. Safinya and J. Raedler (unpublished), Chap. 9.

\bibitem{Mezger2008}
M. Mezger {\it et al.}, Science {\bf 322}, 424 (2008).


\bibitem{Hayes2010}
R. Hayes, G. G. Warr, and R. Atkin, Phys. Chem. Chem. Phys. {\bf 12}, 1709 (2010).

\bibitem{Hoth2014}
J. Hoth, F. Hausen, M. H. M\"user, and R. Bennewitz, J. Phys. Condens. Mat. {\bf 26}, 284110 (2014).

\bibitem{Li2013}
H. Li, F. Endres, and R. Atkin, Phys. Chem. Chem. Phys. {\bf 15}, 14624 (2013).

\bibitem{Perkin2012}
S. Perkin, Phys. Chem. Chem. Phys. {\bf 14}, 5052 (2012).

\bibitem{Kornyshev2008a}
M.V. Fedorov and A. A. Kornyshev, Electrochim. Acta {\bf 53}, 6835 (2008).

\bibitem{Kornyshev2008b}
M. V. Fedorov and A. A. Kornyshev, J. Phys. Chem. B {\bf 112}, 11868 (2008).

\bibitem{Fedorov2013}
K. Kirchner, T. Kirchner, V. Ivani\u{s}t\u{s}ev, and M.V. Fedorov, Electrochim. Acta {\bf 110}, 762 (2013).

\bibitem{Bocquet}
J. Comtet, A. Nigu\`es, V. Kaiser, B. Coasne, L. Bocquet, and A. Siria, Nat. Mater. {\bf 16}, 634, (2017).

\bibitem{Smith2016}
A. M. Smith, A. A. Lee, and S. Perkin, J. Phys. Chem. Lett. {\bf 7}, 2157 (2016).

\bibitem{Underscreening2017}
C. S. Perez-Martinez, A. M. Smith, and S. Perkin, Faraday Discuss. {\bf 199}, 239 (2017).

\bibitem{Israelachvili2013}
M. A. Gebbie, M. Valtiner, X. Banquy, E. T. Fox, W. A. Henderson, and J. N. Israelachvili, Proc. Natl. Acad. Sci. USA {\bf 110}, 9674 (2013).

\bibitem{Gebbie2017}
M. A. Gebbie et al., Chem. Commun. {\bf 53}, 1214 (2017).

\bibitem{Netz2000Beyond}
R. R. Netz and H. Orland, 	Eur. Phys. J. E {\bf 1}, 203 (2000).

\bibitem{Naji2013}
A. Naji, M. Kandu\u c, J. Forsman, and R. Podgornik, J. Chem. Phys. {\bf 139}, 150901 (2013).

\bibitem{Lauw2009}
Y. Lauw, M. D. Horne, T. Rodopoulos, and F. A. M. Leermakers, Phys. Rev. Lett. {\bf 103}, 117801 (2009).

\bibitem{Gavish2016}
N. Gavish and A. Yochelis, J. Phys. Chem. Lett. {\bf 7}, 1121 (2016).

\bibitem{Gavish2017}
N. Gavish, D. Elad, and A. Yochelis, J. Phys. Chem. Lett. {\bf 9}, 36 (2017).

\bibitem{Girotto2017}
M. Girotto, T. Colla, A. P. dos Santos, and Y. Levin, J. Phys. Chem. B {\bf 121}, 6408 (2017).

\bibitem{Tosi1986}
M. Revere and M. P. Tosi, 	Rep. Prog. Phys. {\bf 49}, 1001 (1986).

\bibitem{Zerah1986}
G. Zerah and J. P. Hansen, J. Chem. Phys. {\bf 84}, 2336 (1986).


\bibitem{1D2012a}
V. D\' emery, D. S. Dean, T. C. Hammant, R. R. Horgan, and R. Podgornik, Europhys. Lett., {\bf 97}, 28004 (2012).

\bibitem{1D2012b}
V. D\' emery, D. S. Dean, T. C. Hammant, R. R. Horgan, and R. Podgornik, J. Chem. Phys. {\bf 137}, 064901 (2012).

\bibitem{Bazant}
M. Z. Bazant, B. D. Storey, and A. A. Kornyshev, Phys. Rev. Lett. {\bf 106}, 046102 (2011).

\bibitem{Borukhov1997}
I. Borukhov, D. Andelman, and H. Orland, Phys. Rev. Lett. {\bf 79}, 435 (1997).

\bibitem{Kornyshev1978}
A. A. Kornyshev, A. I. Rubinshtein, and M. A. Vorotyntsev, J. Phys. C Solid State {\bf 11}, 3307 (1978).

\bibitem{Blossey2017}
R. Blossey, A. C. Maggs, and R. Podgornik, Phys. Rev. E {\bf 95}, 060602(R) (2017).

\bibitem{Bjerrum1926}
N. Bjerrum, Kgl. Dan. Vidensk. Selsk. Mat. Fys. Medd. {\bf 7}, 1 (1926).

\bibitem{Abascal1994}
J. L. F. Abascal, F. Bresme, and P. Turq, Mol. Phys. {\bf 81}, 143 (1994).

\bibitem{Levin1996}
Y. Levin and M. E. Fisher, Physica A {\bf 225}, 164 (1996).

\bibitem{Feng2018}
G. Feng, M. Chen, S. Bi, Z. A. H. Goodwin, E. B. Postnikov, N. Brilliantov, M. Urbakh, A. A. Kornyshev, Phys. Rev. X {\bf 9}, 021024 (2019).

\bibitem{VanRoij2009}
J. Zwanikken and R. Van Roij, J. Phys. Condens. Mat. {\bf 21}, 424102 (2009).

\bibitem{LeeAlpha}
A. A. Lee, D. Vella, S. Perkin, and A. Goriely, J. Phys. Chem. Lett. {\bf 6}, 159 (2014).

\bibitem{Kjellander2016}
R. Kjellander, J. Chem. Phys. {\bf 145}, 124503 (2016).

\bibitem{Adar2017}
R. M. Adar, T. Markovich, and D. Andelman, J. Chem. Phys. {\bf 146}, 194904 (2017).

\bibitem{Levy2013}
A. Levy, D. Andelman, and H. Orland, J. Chem. Phys. {\bf 139}, 164909 (2013).

\bibitem{Buyukdagli2013}
S. Buyukdagli and T. Ala-Nissila, Phys. Rev. E {\bf 87}, 063201 (2013).

\bibitem{Frydel2013}
D. Frydel and Y. Levin, J. Chem. Phys. {\bf 138}, 174901 (2013).

\bibitem{Yokoyama1975}
H. Yokoyama and H. Yamatera, B. Chem. Soc. Jpn. {\bf 48}, 1770 (1975).

\bibitem{Ebeling1980}
W. Ebeling and M. Grigo. Annalen der Physik {\bf 492}, 21 (1980).

\bibitem{Podgornik2018}
R. Podgornik, J. Chem. Phys. {\bf 149}, 104701 (2018).

\bibitem{Bopp1996}
P. A. Bopp, A. A. Kornyshev, and G. Sutmann. Phys. Rev. Lett. {\bf 76}, 1280 (1996)

\bibitem{Buyukdagli2013_b}
S. Buyukdagli and T. Ala-Nissila, J. Chem. Phys. {\bf 139} 044907 (2013).

\bibitem{Buyukdagli2014}
S. Buyukdagli and R. Blossey J. Chem. Phys. {\bf 140}, 234903 (2014).

\bibitem{Kjellander2018}
R. Kjellander, J. Chem. Phys. {\bf 148}, 193701 (2018).

\bibitem{Budkov2018}
Y. A. Budkov, J. Phys. Condens. Mat. {\bf 30}, 344001 (2018).

\bibitem{Adar2019}
R. M. Adar, S. A. Safran, H. Diamant, and D. Andelman, Phys. Rev. E {\bf 100}, 042615 (2019).

\bibitem{Supplemental}
See Supplemental Material at [URL will be inserted by publisher] for further details.

\bibitem{Israelachvili1988}
J. N. Israelachvili and P. M. McGuiggan, Science {\bf 241} 795 (1988).

\bibitem{Slavchov1}
R. I. Slavchov and T. Ivanov, J. Chem. Phys. {\bf 140}, 074503 (2014).

\bibitem{Slavchov2}
R. I. Slavchov, J. Phys. Chem. Phys. {\bf 140}, 164510 (2014).

\end{thebibliography}
\end{document}